\documentclass[final,1p,times]{elsarticle} 
\usepackage{graphicx} 
\usepackage{amssymb} 
\usepackage{amsthm} 
\usepackage{lineno} 

\journal{Nuclear Physics A} 
\begin{document} 

\begin{frontmatter} 


\title{Heavy Flavor Tracker (HFT) :  A new inner tracking device at STAR}

\author{J. Bouchet$^{a}$, for the STAR Collaboration}

\address[a]{Kent State University, 
Kent, Ohio, USA}

\begin{abstract} 
The HFT, a new inner tracking detector for STAR, aims to measure the charmed hadron nuclear modification factor as well as their elliptic flow to the low $p_\mathrm{T}$ region ($\sim$0.5 GeV/c) by measuring the displaced vertices of charmed particles.
\end{abstract} 

\end{frontmatter} 

\section{Introduction}
Due to their large masses, heavy flavor ($c$ and $b$) quarks are produced in the early stages of heavy ion collisions \cite{ref1}. 
A precise measurement of heavy flavor production could be achieved by identifying the decay of charmed mesons using direct topological reconstruction and thus disentangling the $c$ and $b$ contributions. The HFT has the necessary resolution for such a measurement that requires high precision. It is the assembly of the existing Silicon Strip Detector (SSD) and 2 new detecting devices: the Intermediate Silicon Tracker (IST) and the PIXEL detector. The PIXEL is composed by 2 layers of monolithic CMOS Active Pixel sensors \cite{ref2} which measure with great accuracy the position of a particle within a few centimeters of the interaction region. These very thin layers 
minimize the multiple coulomb scaterring.
The intermediate tracking system is made by the IST and the existing SSD (Table \ref{tab:fit-res}). The purpose of the IST and the SSD is to link tracks found in the STAR Time Projection Chamber (TPC) to the PIXEL detector. 
\begin{table}[ht]
\caption{Characteristics of each silicon layer of the HFT\label{tab:fit-res}}
\begin{center}
\item[]\begin{tabular}{c c c c c c}
\hline
Detector&Radius&Technology&Si thickness&Hit resolution&Material Budget\\
                &             &                     &                              &R/$\phi-Z$                     & in radiation length $X_\mathrm{0}$  \\
                &(cm)&                     &      ($\mu$m)       &  ($\mu$m - $\mu$m)       &                  \\
\hline
SSD&23&double sided strips&300&30 - 857&1\%\\
IST&14&Si Strip Pad sensors&300&170 -1700&1.2\% \\
PIXEL&2.5, 8&Active Pixels& 50&8.6 - 8.6&0.37\%\\
\hline
\end{tabular}
\end{center}
\end{table}
\section{Simulation details}
Simulations presented in this proceedings were performed using the full STAR geometry package with 10k AuAu HIJING central events at $\sqrt{s_{NN}}$ = 200 GeV embedded with $D^\mathrm{0}$ and $\Lambda_\mathrm{c}$ particles, forced to decay to their hadronic channels ($D^\mathrm{0}$$\rightarrow$$K^\mathrm{-}$$\pi^\mathrm{+}$,  $\Lambda_\mathrm{c}$$\rightarrow$$K^\mathrm{-}$$\pi^\mathrm{+}p$). Their reconstruction efficiencies are based on particle identification of daughter particles provided by the TPC and extended to higher $p_\mathrm{T}$ with the Time of Flight detector (TOF) : $K$-$\pi$ and ($K$+$\pi$)-$p$ separations were done up to $p_\mathrm{T}$ $\leq$ 1.6 GeV/c and $p_\mathrm{T}$ $\leq$ 3 GeV/c, respectively. Topological cuts have been also applied to the $D^\mathrm{0}$ candidates. The effect of {\it out of time} events is included in the PIXEL simulation at a rate corresponding to RHIC-II luminosity.

\section{Estimated $D^\mathrm{0}$ and $\Lambda_\mathrm{c}$ reconstruction performances}
\begin{figure}[!h]
\begin{center}
\includegraphics[width=0.28\textwidth]{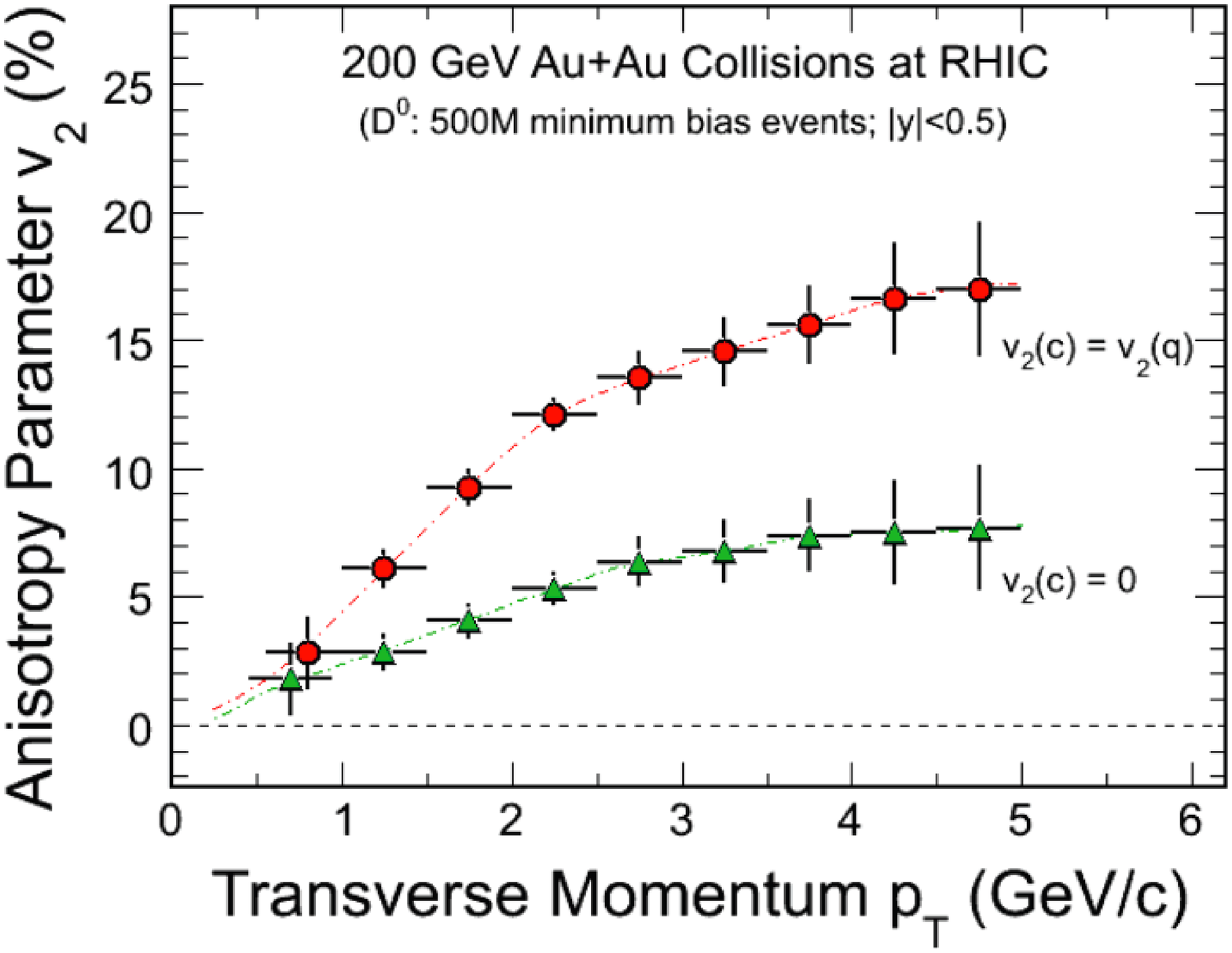}
\includegraphics[width=0.28\textwidth]{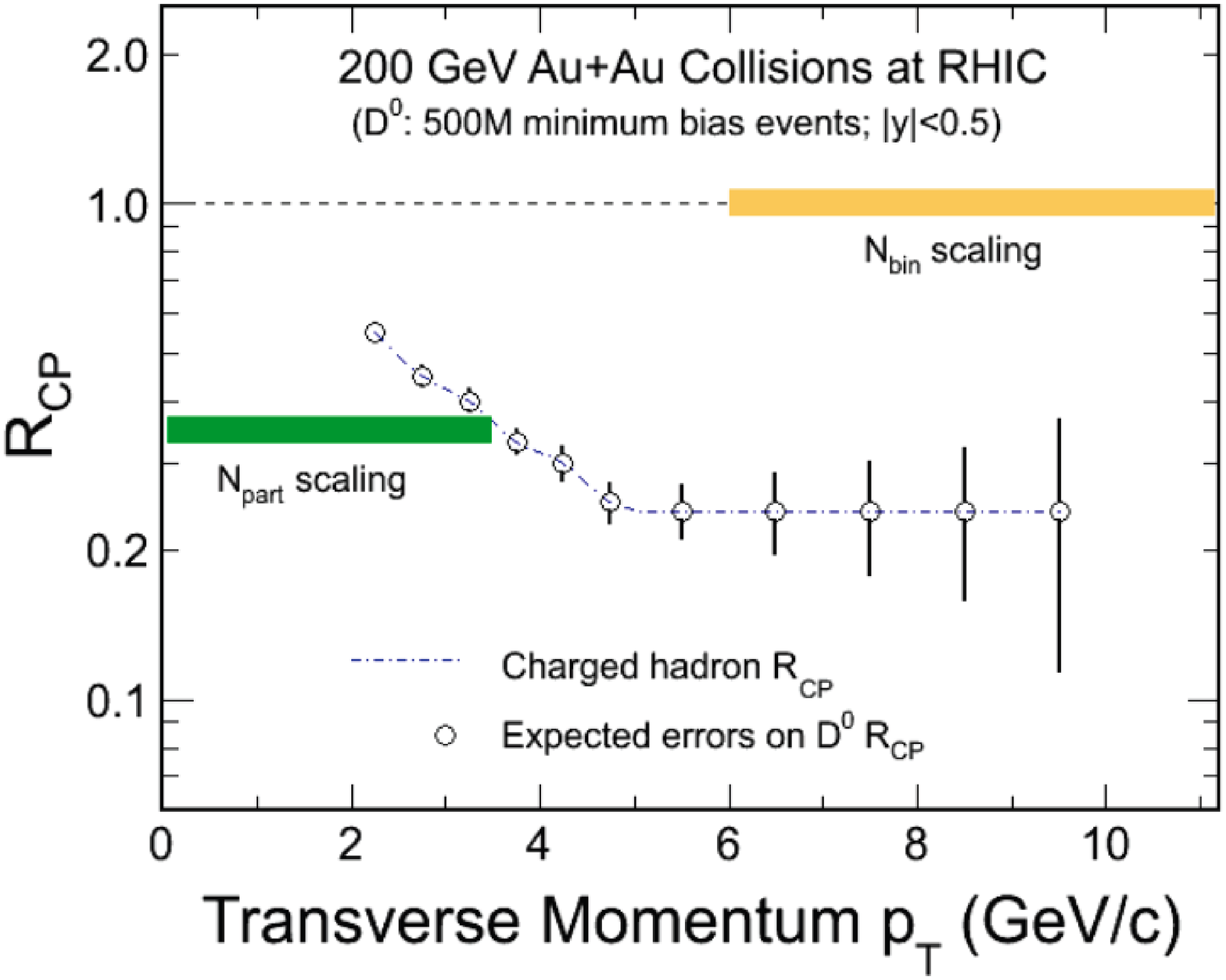}
\includegraphics[width=0.28\textwidth]{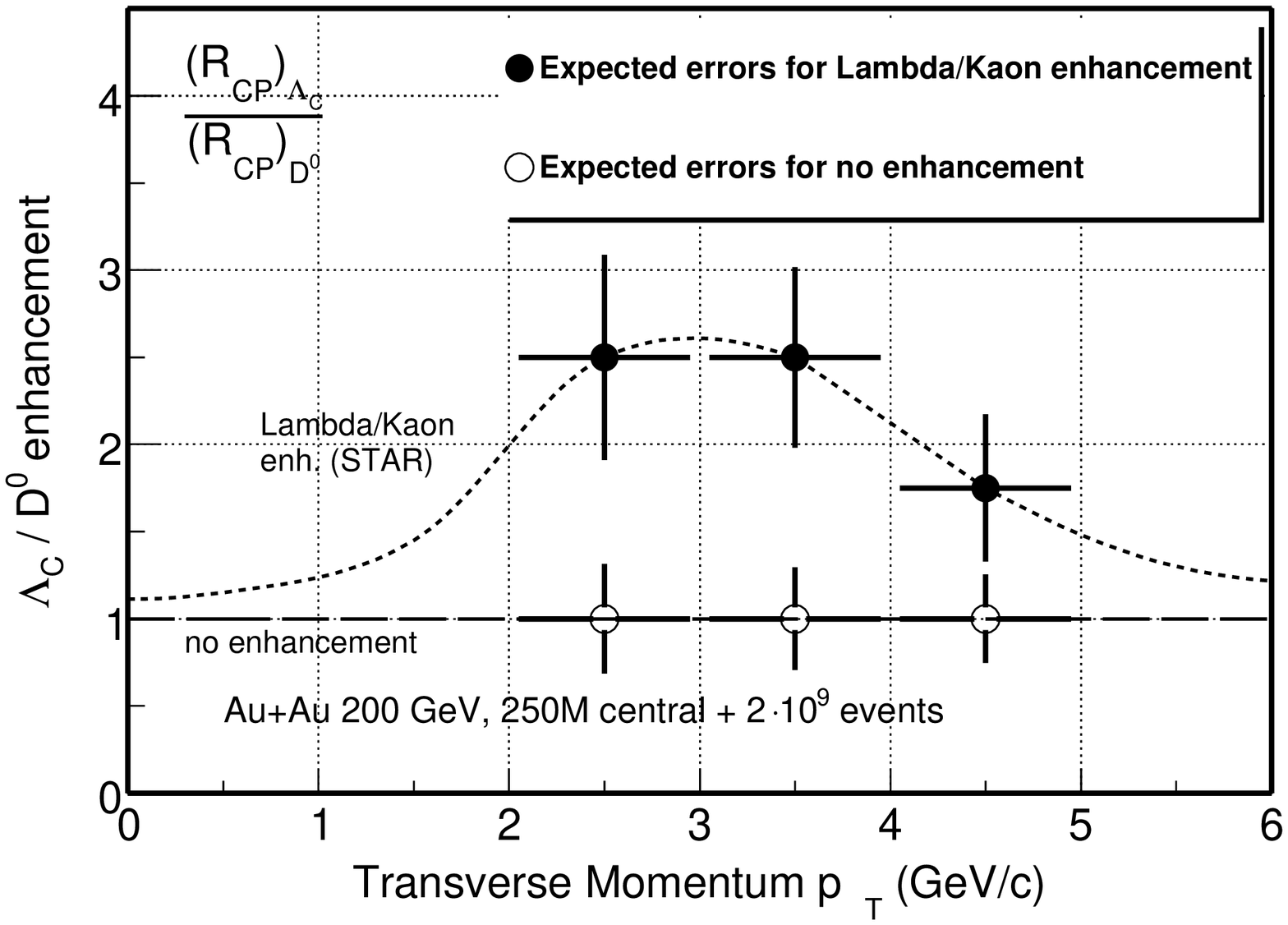}
\end{center}
\caption{Projections of key measurements with HFT\label{fig:keytest}}
\end{figure} 
Figures \ref{fig:keytest} show the statistical error projections for the key measurements with the HFT in 500 M minimum bias AuAu collisions. Fig. \ref{fig:keytest}(left) is the flow parameter $v_{2}$, shown for two extreme scenarios [charm quark flow equal to light quark flow (red circles) and charm quark does not flow (green triangles)]. The HFT will be able to distinguish with great accuracy these 2 cases. Figure \ref{fig:keytest}(middle) shows the suppression factor $R_\mathrm{CP}$ of $D^\mathrm{0}$: the HFT will be able to measure it directly for $p_\mathrm{T}$ $\leq$ 10 GeV/c via the hadronic channel thus avoiding the indirect method using non-photonic electrons. A measurement of $\Lambda_\mathrm{c}$ is important to perform since the $\Lambda_\mathrm{c}$/$D^\mathrm{0}$ ratio may be enhanced, indicating a similar pattern to the baryon/meson ratio involving light quarks in the intermediate $p_\mathrm{T}$ region \cite{ref3}. Two scenarios are investigated for the $\Lambda_\mathrm{c}$/$D^\mathrm{0}$ ratio \cite{ref4} : no enhancement and same enhancement as $\Lambda$/$K_\mathrm{s}^\mathrm{0}$.
We see from Fig. \ref{fig:keytest}(right) that the statistical errors are sufficiently small, making a measurement of baryon/meson ratio in charm sector with good precision in heavy ion collisions.
 \section{Summary}
The HFT, by using low mass CMOS sensors, will be able to directly reconstruct charm hadrons over a large momentum range and, thus, study flow and energy loss of heavy flavor particles. Several physics capabilities such as baryon/meson ratio in the charm sector have been studied. 


\begin{thebibliography}{00} 
\bibitem{ref1}Z. Lin and M. Gyulassy, {\it Phys. Rev. C} {\bf77} (1996) 1222.
\bibitem{ref2}E. Anderssen et al., {\it A Heavy Flavor Tracker for STAR} (http://www.osti.gov/bridge/servlets/purl/939892-be12Up/939892.pdf).
\bibitem{ref3}S. H. Lee et al., {\it Phys.\ Rev.\ Lett.} {\bf 100} (2008) 222301.
\bibitem{ref4}J. Kapitan, {\it Eur. Phys. J. C} {\bf62} (2009) 217-221.
\end{thebibliography}
\end{document}